\documentclass[twocolumn,showpacs]{revtex4}

\usepackage{graphicx}
\usepackage{amsmath}
\usepackage{amssymb}
\usepackage{epsfig}

\makeatletter
\def\btt#1{\texttt{\@backslashchar#1}}
\DeclareRobustCommand\bblash{\btt{\@backslashchar}}
\makeatother

\begin{document}

\title[Short Title]{Bang-Bang Operations from a Geometric Perspective}

\author{Mark S. Byrd}
\email{mbyrd@chem.utoronto.ca}
\author{Daniel A. Lidar}
\email{dlidar@chem.utoronto.ca}

\affiliation{Chemical Physics Theory Group, University of Toronto, 
80 St. George Street, Toronto, Ontario M5S 3H6, Canada }

\date{\today}

\begin{abstract}
Strong, fast pulses, called ``bang-bang'' controls can be 
used to eliminate the effects of 
system-environment interactions.  This method for 
preventing errors in quantum information processors is treated here in 
a geometric setting which leads to an intuitive perspective.  Using 
this geometric description, we clarify the notion of group symmetrization as an 
averaging technique, and provide a geometric picture for evaluating 
errors due to imperfect bang-bang controls.  
This will provide additional support for the usefulness of such controls 
as a means for providing more reliable quantum information processing.  
\end{abstract}

\maketitle



\def\Bid{{\mathchoice {\rm {1\mskip-4.5mu l}} {\rm
{1\mskip-4.5mu l}} {\rm {1\mskip-3.8mu l}} {\rm {1\mskip-4.3mu l}}}}



\section{Introduction}

Recently, controlling the evolution of a system by using 
strong, short pulses has been introduced as a new means for 
quantum error correction/prevention 
\cite{Viola:98,Vitali:99,Zanardi:98b,Duan:98e}.  
These operations have been 
termed ``bang-bang'' (BB) pulses \cite{Viola:98}
(a name derived from classical control theory 
\cite{Jurdjevic:book}), ``parity 
kicks'' \cite{Vitali:99,Vitali:01} (for the special case of 
a sign changing operation), decoupling operations 
\cite{Viola:99a,Viola:00a} (since they 
can serve to decouple the 
system from environmental degrees of freedom), and symmetrization 
procedures \cite{Zanardi:98b} 
(which are associated with a group symmetrization/averaging).  
The advantage they have over the active and passive 
error correction procedures associated with 
quantum error correcting codes (QECCs) 
(see \cite{Shor:95,Gottesman:97a,Steane:99} and references therein)
and decoherence free subspaces (DFSs)
(or noiseless subsystems; see \cite{Kempe:00,Knill:99a,Zanardi:99d} 
and references therein) is the use of external pulses 
rather than requiring 
several physical qubits to encode one logical qubit.  
Since today's experiments use $<10$ qubits, 
this may, for the time being, make BB controls 
a method of choice for small-scale quantum 
computer implementations.  
However, it is clear that the time constraints imposed 
by bang-bang operations on the system are great 
\cite{Viola:99a} and may not be 
practical for eliminating noise altogether.  Even if this 
technique cannot completely eliminate the noise, it can still 
be used to reduce noise \cite{Viola:98,Duan:98e,Vitali:01}.  
This is important for possibly reducing 
error rates and thus extending computing time and/or the utility of 
QECCs and/or DFSs.  

In order to take full advantage of the BB technique, 
the symmetrization operations and their effects 
must be made clear so that the benefit from the implementation 
can be readily determined.  
The work put forth here will aid in the analysis of 
the results of BB operations by providing an explicit geometric representation for the group-algebraic elements describing 
such interactions.  
This geometric picture also has the advantage of clarifying 
the error between a desired and a 
modified evolution.  In addition, 
for two-state systems (qubits), we recover a 
familiar Bloch sphere representation and this provides us 
with an intuitive understanding of BB/symmetrization 
operations.


\section{Decoupling by Symmetrization}

The process of decoupling by symmetrization, counteracts decoherence 
by applying sequences of frequent pulses 
\cite{Vitali:99,Viola:98,Duan:98e}.  The time scales are crucial: roughly
speaking, one needs to perform a complete cycle of symmetrization operations
in a time shorter than the bath correlation time.  An elegant group
theoretical treatment shows that if the applied pulses are unitary
transformations forming a finite-dimensional 
group, then the application of that series of pulses amounts 
to an average (symmetrization) over this group 
\cite{Viola:99,Viola:99a,Viola:00a,Zanardi:98b,Zanardi:99d}. 
We briefly review this theory.

The general evolution of a system and a bath coupled to it can be
written in the form 
\begin{equation}
H = H_S \otimes \Bid_B + \Bid_S \otimes H_B + \sum_\gamma S_\gamma 
\otimes B_\gamma,
\label{eq:H}
\end{equation}
where $H_S$ acts on the system alone, $H_B$ acts on the bath 
alone, and $H_I\equiv \sum_\gamma S_\gamma \otimes B_\gamma$ 
is the interaction part of the Hamiltonian composed of traceless
operators $S_\gamma$ ($B_\gamma$) which act on the system (bath).  The objective of 
the BB procedure is to modify this evolution.

A set of symmetrization or BB 
operations can be chosen such that they form a 
discrete (finite order) subgroup of
the full unitary group of operations on the Hilbert space of the system.
Denote this subgroup $\mathcal{G}$ and its elements 
$g_{k}$, $k=0,1,...,|\mathcal{G}|-1$, where $|\mathcal{G}|$ is the order of the group. The cycle
time is $T_{c}=|\mathcal{G}|\Delta t$, where $|\mathcal{G}|$ is now
also the number of
symmetrization operations, and $\Delta t$ is the time that the system
evolves freely between operations under $U_{0}=\exp(-iHt)$.  
The symmetrized evolution is given by 
\begin{equation}
U(T_{c})=\prod_{k=0}^{|\mathcal{G}|-1}g_{k}^{\dagger }U_{0}(\Delta t)g_{k}\equiv
e^{iH_{eff}T_{c}}.  \label{exactevol}
\end{equation}
$H_{eff}$ denotes the resulting effective Hamiltonian.  Since the 
approximation requires very strong, short pulses to be 
implemented in a sequence, 
they have been termed bang-bang (BB) operations (we will use 
symmetrization and BB operations interchangeably).  In this (BB) limit  
\begin{equation}
H\mapsto H_{eff}=\frac{1}{|\mathcal{G}|}\sum_{k=0}^{|\mathcal{G}|-1}g_{k}^{\dagger }Hg_{k}\equiv \Pi
_{\mathcal{G}}(H),  \label{Heff}
\end{equation}
where $H_{eff}$ is the desired Hamiltonian (without noise).  The map $\Pi _{%
\mathcal{G}}$ is the projector into the centralizer, $Z(\mathcal{G})$,
defined as 
\begin{equation}
Z(\mathcal{G})=\{X|\;[X,g_{k}]=0,\;\forall g_{k}\in \mathcal{G}\}. 
\end{equation}
It is clear that $\Pi _{\mathcal{G}}$ commutes with all $g_{k}$ so that, if
our group is generated by $\{{%
\mathchoice {\rm {1\mskip-4.5mu l}} {\rm
{1\mskip-4.5mu l}} {\rm {1\mskip-3.8mu l}} {\rm {1\mskip-4.3mu l}}}%
,H_{S},S_{\gamma }\}$, the system is effectively decoupled 
from its environment.  The \emph{control algebra} is the algebra
generated by the set $\{g_{k}\}$.  Even if the symmetrization is performed
under less than ideal conditions, it can still reduce the noise in the
system \cite{Viola:98,Duan:98e,Vitali:01}.


\section{Geometric Interpretation of the Effect of BB Operations}

\label{geom}

Now consider a set of unitary operators 
$\{U_k\}$, $U_0 \equiv \Bid_S$, as an explicit
realization of the subgroup ${\cal G}$ and choice of our 
set of BB operations. Then the following condition must be satisfied
for an evolution generated by the effective Hamiltonian:
\begin{equation}
\label{noteequ}
H_{eff}=\frac{1}{|\mathcal{G}|}\sum_{k=0}^{|\mathcal{G}|-1} U_k^\dagger H U_k.
\end{equation}
Note that $H_{eff}=0$ is the case of \emph{storage}. Considering
Eq.~(\ref{eq:H}), we can always include the terms $H_S \otimes \Bid_B$ and
$\Bid_S \otimes H_B$ in $H_I$. We do not include the identity
component $\Bid_S \otimes \Bid_B$ since it only gives rise to an
overall phase. Thus $H$ and $H_{eff}$ are traceless. Let us now introduce
$N \equiv n^2-1$ traceless, Hermitian generators $\{ \lambda_i \}_{i=1}^N$ of
$SU(n)$. These generators are
closed under commutation and span the space of 
traceless Hermitian matrices. For $SU(2)$, 
the Pauli matrices are commonly used; for $SU(3)$, the 
Gell-Mann matrices, and for higher dimensions, one may use
a direct generalization of the Gell-Mann 
matrices. For dimensions that are a power of two it is often
convenient to use the Pauli group (tensor products of Pauli
matrices). The $\{\lambda_i\}$ satisfy trace-orthogonality,
\begin{equation}
\mbox{Tr}(\lambda_i \lambda_j) = M \delta_{ij},
\label{eq:tr}
\end{equation}
where $M$ is a normalization constant (often taken to be $2$ for 
Lie algebras or 
$n$ for $n\times n$ matrices).  
Expanding the system operators in terms of the $\{\lambda_i\}$ yields:
\begin{equation}
S_\gamma = \sum_i a_{i \gamma} \lambda_i
\end{equation}
where the expansion coefficients are
\begin{equation}
a_{i \gamma} = \frac{1}{M} \mbox{Tr}(\lambda_i S_\gamma).
\label{eq:ai}
\end{equation}
Using this, $H$ can be written as as follows:
\begin{equation}
\label{Had}
H = \sum_\gamma S_\gamma \otimes B_\gamma = \sum_\gamma
\sum_{i=1}^{N}a_{i \gamma} \lambda_i \otimes B_\gamma \equiv
\sum_\gamma (\vec{a}_\gamma \cdot \vec{\lambda}) \otimes B_\gamma .
\end{equation}
Here $\vec{a}_\gamma$ and $\vec{\lambda}$ are vectors of length $N$.
In this representation, used extensively in \cite{Mahler:book}, an
$n\times n$ Hamiltonian, $H$, is a vector with coordinates
$\vec{a}_\gamma$ for each error $\gamma$
in an $N$-dimensional vector space spanned by the $\{ \lambda_i \}$ as
basis vectors, with ordinary vector addition and scalar multiplication. 

As is well-known, there is a homomorphic mapping between the Lie
groups $SU(2)$ and $SO(3)$ \cite{Tinkham:book}. This mapping is
generalized as follows for $SU(n)$ and a subgroup of the 
rotation group $SO(N)$:
\begin{equation}
U_k^\dagger \lambda_i U_k = \sum_{j=1}^{N} R_{ij}^{(k)} \lambda_j,
\label{eq:R}
\end{equation}
where the matrix $R^{(k)} \in SO(N)$, 
the adjoint representation of $SU(n)$.

The BB operation [Eq.~(\ref{noteequ})] may now be viewed as 
a weighted sum of rotations of the (adjoint) vectors 
$\vec{a}_\gamma$. To see this, first let
\begin{equation}
\vec{a}_\gamma^{(k)} = R^{(k)} \vec{a}_\gamma .
\end{equation}
This represents the rotation by $R^{(k)}$ of the coordinate vector
$\vec{a}_\gamma$. Next average over all rotations:
\begin{equation}
\vec{a}_\gamma^{\prime} = \frac{1}{|\mathcal{G}|} \sum_{k=0}^{|\mathcal{G}|-1} \vec{a}_\gamma^{(k)} .
\end{equation}
Finally, note that the effective Hamiltonian, after the BB operations,
can be rewritten as:
\begin{equation}
\label{Hgeoeff}
H_{eff} = \frac{1}{|\mathcal{G}|}\sum_{k=0}^{|\mathcal{G}|-1} U_k^\dagger H U_k = 
               \sum_{\gamma} (\vec{a}^\prime_\gamma \cdot
               \vec{\lambda}) \otimes B_\gamma .
\end{equation} 
Eq.~(\ref{Hgeoeff}) [compare to Eq.~(\ref{Had})] is our 
desired geometric representation of BB
operations. Their effect is to simply transform, for each error
$\gamma$, the coordinates $\vec{a}_\gamma$ to
$\vec{a}^\prime_\gamma$. It is simplest to interpret 
this in the case of storage, where
we seek BB operations such that $H_{eff} = 0$. Since the errors can 
be decomposed in the linearly independent basis set indexed by $\gamma$, 
each term $\vec{a}^\prime_\gamma \cdot \vec{\lambda}$ must
vanish separately. Furthermore, since the $\lambda_i$ are independent
this can only be satisfied if $\vec{a}^\prime_\gamma = \vec{0}$ for
each $\gamma$. This means that 
\begin{equation}
\vec{a}_\gamma^{\prime} = \left( \frac{1}{|\mathcal{G}|} \sum_k R^{(k)} \right)
\vec{a}_\gamma = \vec{0},
\end{equation}
i.e., the sum of all rotations applied to the original coordinate
vector $\vec{a}_\gamma$ must vanish.

Similarly, to obtain a 
modified evolution corresponding to a 
target Hamiltonian $H_{eff}^t = \sum_{\gamma} (\vec{a}^t_\gamma \cdot
\vec{\lambda}) \otimes B_\gamma$,  we require the weighted sum of 
rotations applied to the original coordinate vector to be equal to the
corresponding target coordinate vector $\vec{a}_\gamma^{t}$. I.e., for $H_{eff} \neq 0$, the following 
condition should be satisfied to obtain the desired 
evolution:
\begin{equation}
\label{hgeff}
\vec{a}_\gamma^{\prime} = \vec{a}_\gamma^{t}
\end{equation}
This may require a combination of switching strategies 
for the BB pulses \cite{Viola:00a}.

It should be noted that our geometrical picture is an explicit representation of a 
subset of the 
group algebra $\mathbb{C}{\cal G}$ 
using the set of traceless 
Hermitian matrices and the identity as 
the basis.  When the coefficients of 
the adjoint vector are real, the resulting matrix $H_{eff}$ is Hermitian.  
When they are complex, the resulting matrix is not Hermitian 
and the evolution is not unitary but may still be 
treated empirically \cite{Byrd/Lidar:Roch}.


\section{Errors}

The picture developed above also gives an intuitive way in which to 
evaluate the error that remains in the system 
evolution after the application of the BB pulses.  
Let $\vec{a}_\gamma^{t}$ be the coordinates vector 
corresponding to the desired Hamiltonian evolution 
and $\vec{a}_\gamma^{\prime}$ the actual vector after BB 
operations.  Then $\vec{a}_\gamma^{\prime}$ corresponds to the 
effective Hamiltonian, Eq.~(\ref{Hgeoeff}) 
(and may be determined using 
quantum process tomography, see \cite{Byrd/Lidar:Roch} and 
references therein). The error vector 
$\vec{e}$ is given by their difference in the $n^2$-dimensional 
vector space where our geometric picture holds:
\begin{equation}
\label{errorvec}
\vec{e} = \vec{a}_\gamma^{\prime} - \vec{a}_\gamma^{t}
\end{equation}
The vector $\vec{e}$ gives us the 
magnitude and direction of the error ({\it i.e.}, the basis elements
$\lambda_i$ give the type of error, {\it e.g.}, bit-flip 
and/or phase-flip).  Now consider the magnitude of this error, 
\begin{equation}
d(\vec{a}_\gamma^{\prime}, \vec{a}_\gamma^{t}) 
                = (\vec{e}^{\,*}\cdot \vec{e})^{1/2}
\label{eq:d}
\end{equation}
(in the case of Hamiltonian evolution there is no need 
for complex conjugation).
This is the Euclidean distance between the two vectors in the adjoint 
representation space.  For two two-state density matrices, 
it is proportional to 
the Euclidean distance between the two Bloch vectors,
as is the trace distance. In general, computing 
$d(\vec{a}_\gamma^{\prime}, \vec{a}_\gamma^{t})$ is
more manageable than other measures of distance ({\it e.g.}, fidelity
\cite{Nielsen:book}), since it does not require diagonalization.  

In the case of imperfect BB operations, the goal is to minimize 
the distance $d$.
For the purposes of optimization, note that
\begin{equation}
d(\vec{a}_\gamma^{\prime},\vec{a}_\gamma^{t}) = (\vec{e}\cdot \vec{e})^{1/2} 
           = ((a_\gamma^{\prime})^2 + (a_\gamma^{t})^2 
             - 2M \vec{a}_\gamma \cdot \vec{a}_\gamma^{t})^{1/2},
\end{equation}
whereas the ordinary trace distance (Hilbert-Schmidt norm) between
$(\vec{a}_\gamma^{\prime}, \vec{a}_\gamma^{t})$ gives 
[using Eq.~(\ref{eq:tr})]
\begin{equation}
\label{HSnorm}
\mbox{Tr}[(\vec{a}_\gamma\cdot \vec{\lambda})
( \vec{a}_\gamma^t\cdot \vec{\lambda})]
= M\vec{a}_\gamma\cdot \vec{a}_\gamma^t.
\end{equation}
So minimizing $d$ is equivalent to 
maximizing 
$(\vec{a}_\gamma^{\prime}\cdot \vec{a}_\gamma^{t})$.  
The advantage of using $d$ is an intuitive 
one since the error vector simply describes a Euclidean 
vector (in the adjoint representation space).

For obtaining a desired unitary evolution, note that 
trace-norm distance for matrices, $U$ and $V$ is defined by 
\begin{equation}
d_u(U,V) = \sqrt{1-(1/n)\mbox{Re}[\mbox{Tr}(U^\dagger V)]},
\end{equation}
where $U$, and $V$ are $n\times n$ matrices.  For 
BB controls a short-time approximation is relevant.  
For the case of unitary evolution, the 
two measures, Euclidean distance and trace-norm for matrices, 
are equivalent.  Approximating $U$ (desired evolution) and $V$ (actual
evolution) by 
$\Bid -iHt = \Bid-it\sum_i a_{i\gamma}^t\lambda_i$ 
and $\Bid -iH^\prime t = \Bid-it\sum_i a_{i\gamma_i}^\prime \lambda_i$ respectively,  
\begin{eqnarray}
d(U,V) &\approx& \sqrt{1-(1/n)\mbox{Re}[\mbox{Tr}(\Bid + HH^\prime t^2)]} 
\nonumber \\
&\propto& \sqrt{\vec{a}_\gamma^\prime \cdot \vec{a}_\gamma^t}
\end{eqnarray}
is an $O(t)$ approximation to the unitary evolution.  This is 
equivalent to Eq.~(\ref{eq:d}).


\section{Examples}

We now discuss the example of storing a single qubit. In this case our
geometrical picture can be cast in the
familiar Bloch sphere representation.

Consider the noisy evolution of a stored qubit.  Suppose 
the evolution of the qubit is governed by the Hamiltonian 
\begin{equation}
H = \sum_\gamma\sum_{i=1}^3 a_{i \gamma} \sigma_i\otimes B_\gamma,
\end{equation}
where the $a_{i \gamma}$ are real coefficients, $B_\gamma$ are bath operators, 
and the $\sigma_i$ are the Pauli matrices with the 
usual identification $\sigma_1 =\sigma_x$, $\sigma_2 =\sigma_y$, 
$\sigma_3 =\sigma_z$.
(As above, the identity component is neglected.)  
For faithful storage, a set of BB operations $\{ U_k \}$ should serve 
to eliminate this Hamiltonian.  Under such controls, the 
evolution is described by 
\begin{equation}
\label{Hgeoeffsu2}
H_{eff} = \frac{1}{|\mathcal{G}|}\sum_{ik\gamma} U_k^\dagger 
             [(\vec{a}_{\gamma} \cdot \vec{\sigma}) \otimes B_\gamma]U_k 
         = \sum_{\gamma} (\vec{a}_{\gamma}^\prime \cdot \vec{\sigma}) \otimes B_\gamma,
\end{equation}
where $U_k\in SU(2)$ and $R^{(k)} \in SO(3)$, and
\begin{equation}
\vec{a}_{\gamma}^\prime = \frac{1}{|\mathcal{G}|} \sum_{k=0}^{|\mathcal{G}|-1} R^{(k)} \vec{a}_{\gamma}.
\end{equation}
In the subsections below we consider different choices of the subgroup
$\mathcal{G}$. These equations then describe a sum of vectors on the Bloch 
sphere.  The mapping from the unitary matrices $U_k$ to the 
rotation matrices $R^{(k)}$ is given by 
\begin{eqnarray}
U^\dagger \sigma_i U &=& 
e^{i\sigma_3 \alpha/2}e^{i\sigma_2 \beta/2}e^{i\sigma_3\gamma/2} \sigma_i 
e^{-i\sigma_3\gamma/2} e^{-i\sigma_2 \beta/2} e^{-i\sigma_3 \alpha/2} 
         \nonumber \\
                    &=& R_{ij}\sigma_j.
\end{eqnarray}
\begin{widetext}
Explicitly, the rotation matrix is given by
\begin{equation}
R = 
\label{fullR}
\left[
\begin{array}{ccc}
\cos(\alpha) \cos(\beta) \cos(\gamma) 
& -\sin(\alpha) \cos(\beta) \cos(\gamma)
& \sin(\beta) \cos(\gamma) \\
-\sin(\alpha)\sin(\gamma) 
& -\cos(\alpha)\sin(\gamma)  
& \phantom{0} \\
 & & \\
\cos(\alpha)\cos(\beta)\sin(\gamma)  
& -\sin(\alpha) \cos(\beta) \sin(\gamma)
& \sin(\beta)\sin(\gamma) \\
+\sin(\alpha) \cos(\gamma) 
& +\cos(\alpha)\cos(\gamma)
&  \\
& & \\
-\cos(\alpha)\sin(\beta) 
& \sin(\alpha)\sin(\beta)
&\cos(\beta)
\end{array}
\right].
\end{equation}
\end{widetext}
Alternatively, one may use 
\begin{equation}
\exp(i(\theta/2)\hat{n}\cdot\vec{\sigma})\vec{x}\cdot \vec{\sigma} 
\exp(-i(\theta/2)\hat{n}\cdot\vec{\sigma}) 
 = \vec{x}^\prime\cdot \vec{\sigma},
\end{equation}
where 
\begin{equation}
\vec{x}^\prime = (\hat{n}\cdot \vec{x})\hat{n} + [(\hat{n}\times \vec{x})\times \hat{n}]\cos(\theta) + [\hat{n}\times \vec{x}]\sin(\theta).
\end{equation}
The correspondence between the unitary and orthogonal groups 
is made by 
\begin{eqnarray}
\label{Rn-un}
\sum_jx_j^\prime \sigma_j 
       &=&\sum_{i,j} x_i[R_{\hat{n}}(\theta)]_{ij}\sigma_j \nonumber \\
       &=& 
U(\hat{n},\theta/2)\left(\sum_ix_i \sigma_i\right)U^\dagger(\hat{n},\theta/2),
\end{eqnarray}
where $R_{\hat{n}}(\theta)$ is a rotation by $\theta$ about the axis
$\hat{n}$. Although this notation is more compact, the Euler angle 
parameterization of $SU(3)$ and $SU(4)$ have been given 
\cite{Byrd:98,Byrd:Erratum,Tilma:tbp}.


\subsection{Storing One Qubit}
\label{ex:1qubit}

\begin{figure}
\begin{picture}(100,110)(10,20)
\put(50,62){\vector(0,1){50}}
\put(50,62){\vector(1,0){46}}

\put(50,62){\vector(0,-1){50}}
\put(50,62){\vector(-1,0){46}}

\put(-20,68){{\small{$R^{(1)}\vec{a}=\!-\vec{a}$}} }
\put(80,68){$\vec{a}$ }
\put(100,60){$\sigma_3$}

\put(80,62){\circle*{2}}
\put(20,62){\circle*{2}}

\end{picture}


\caption{Addition of adjoint vectors on the Bloch sphere 
corresponding to a single pulse application.  The vertical 
line represents zero $\sigma_3$ component.}
\label{pirotation}

\end{figure}
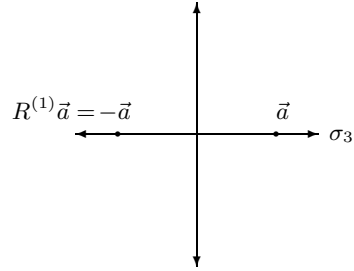

To be specific, consider an unwanted pure dephasing interaction 
described by the Hamiltonian 
\begin{equation}
\label{Hz}
H = g\sigma_3\otimes B.
\end{equation}
Using Eq.~(\ref{eq:ai}) we find that the coordinate (adjoint) 
representation of this Hamiltonian is
\begin{equation}
a_i = \frac{g}{2} \mbox{Tr} (\sigma_i \sigma_3) = g \delta_{i3} ,
\end{equation}
i.e., $\vec{a} = (0,0,g)$.  This dephasing could be corrected through
the use of a single BB operation $U_1= \exp(-i\sigma_1 \pi/2) =
-i\sigma_1$, so that
\begin{equation}
U_1^\dagger \sigma_3 U_1 =  -\sigma_3 .
\end{equation}
For the geometric picture, using $\beta = \pi,\;\alpha=\gamma=0$ 
in Eq.~(\ref{fullR}):
\begin{equation}
R^{(1)}= \left(\begin{array}{ccc}
   -1 &  0  & 0 \\ 
   0 & 1  & 0 \\
   0 &  0  & -1
  \end{array}\right) ,
\end{equation}
which inverts the adjoint vector $\vec{a}$. This is shown 
schematically in Fig. \ref{pirotation}.  It is simple to check that, 
as required, $H_{eff} = 0$. This example uses the lowest dimensional
finite order group $C_2$.

\begin{figure}
\begin{picture}(100,110)(10,20)
\put(50,65){\vector(0,1){50}}
\put(50,65){\vector(1,0){46}}

\put(50,65){\vector(0,-1){50}}
\put(50,65){\vector(-1,0){46}}

\put(50,65){\vector(-1,2){16}}
\put(50,65){\vector(-1,-2){16}}

\put(45,120){$\sigma_2$}
\put(100,63){$\sigma_3$}

\put(7,100){$R^{(1)}\vec{a}$}
\put(-5,26){$(R^{(1)})^2\vec{a}$}
\put(82,69){$\vec{a}$}

\put(84,65){\circle*{2}}

\end{picture}

\caption{Addition of adjoint vectors on the Bloch sphere corresponding 
to BB operations.}
\label{3rdpirotation}

\end{figure}
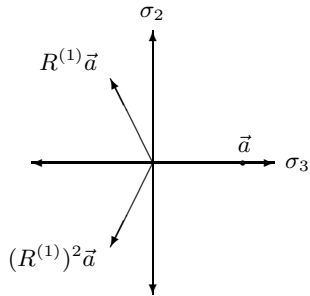

Now, let us use our geometric picture to derive another class of BB
operations for pure dephasing on a single qubit. Clearly, the point is
to find a set of rotations of $\vec{a}$ which when added sum up to
zero. The next example is the group $C_3$, which consists of rotating
$\vec{a}$ by $2\pi/3$ and $-2\pi/3$ about a fixed axis, 
i.e., uses two non-trivial BB
operations.  This is depicted in Fig. \ref{3rdpirotation} where 
we have chosen $\sigma_1$ as the fixed axis.  
The set of rotation matrices that accomplish this are 
$R^{(1)}=R_{\hat{x}}(2\pi/3)$ and $R^{(2)}=R_{\hat{x}}(-2\pi/3)=(R^{(1)})^2$, where 
\begin{eqnarray} 
R_{\hat{x}}(\theta) &=&  \left(\begin{array}{ccc}
 	  1 &  0  & 0 \\ 
  	  0 & \cos(\theta)  & \sin(\theta) \\
   	  0 &   -\sin(\theta) & \cos(\theta)
          \end{array}\right)     
\end{eqnarray}
and the corresponding unitary transformations are 
$U_k = \exp(-i\sigma_1 (\pm\pi/3))$, $k=1,2$.

A large number of 
BB operations is undesirable due to time constraints so 
that large sets become successively more difficult to 
implement effectively.  However, 
depending on the symmetries of the Hamiltonian and the 
experimentally available BB operations, using a larger 
number may be advantageous.  

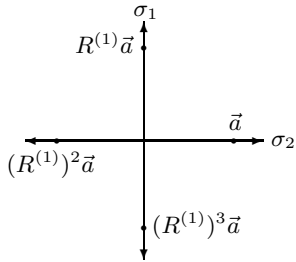
\begin{figure}
\begin{picture}(100,90)(10,20)
\put(50,62){\vector(0,1){45}}
\put(50,62){\vector(1,0){45}}

\put(50,62){\vector(0,-1){45}}
\put(50,62){\vector(-1,0){45}}

\put(46,109){$\sigma_1$}
\put(98,60){$\sigma_2$}

\put(84,62){\circle*{2}}\put(82,66){$\vec{a}$}
\put(50,97){\circle*{2}}\put(24,95){$R^{(1)}\vec{a}$}
\put(17,62){\circle*{2}}\put(-2,50){$(R^{(1)})^2\vec{a}$}
\put(50,29){\circle*{2}}\put(53,27){$(R^{(1)})^3\vec{a}$}

\end{picture}


\caption{The application of the cyclic group of order 4. Here
$R^{(1)}=R_{\hat{x}}(\pi/4)$.}
\label{C4}

\end{figure}

Let us consider the next higher order set of BB operations.  This 
will be a set with 4 elements.  The subgroup condition requires 
a set forming either the cyclic group of order 4 or the 
so-called Vierergruppe since these are the only two 
groups of order 4 \cite{Tinkham:book}.  An example of the 
cyclic group of order 4 would be the four-fold rotations about 
a single axis, ({\it e.g.}, $\pi/2$ around the $\sigma_1$ axis, as in 
Fig. \ref{C4}).  An example of the the other fourth order group 
is the set of rotations by $\pi$ about three orthogonal symmetry 
axes.

\begin{center}
\begin{figure}[hhh]
\unitlength1mm
\begin{picture}(80,80)
\put(10,0){
\resizebox{7cm}{!}{\includegraphics{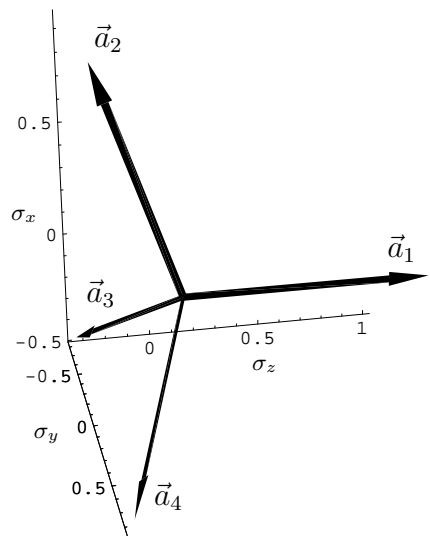}}}
\put(14,16){$\sigma_y$}
\put(11,45){$\sigma_x$}
\put(43,25){$\sigma_z$}
\put(61,40){{\large $\vec{a}_1$}}
\put(21,34){{\large $\vec{a}_3$}}
\put(22,68){{\large $\vec{a}_2$}}
\put(30,7){{\large $\vec{a}_4$}}

\end{picture}
\caption{Rotations of the Hamiltonian to vertices of the
Tetrahedron. The rotated vectors are $\vec{a}_2 =
R_{\hat{y}}(\theta)$, $\vec{a}_3 = R_{\hat{a}_2}(2\pi/3)$, 
$\vec{a}_4 = R_{\hat{a}_2}(-2\pi/3)$.}
\label{tverts}
\end{figure}
\end{center}

Note that the set of vectors pointing to the vertices of a 
tetrahedron also will sum to zero and thus form a set of 
adjoint vectors, representing BB modified 
Hamiltonians, that will produce the 
desired decoupling effect, the elimination of 
the interaction Hamiltonian Eq. (\ref{Hz}).  This set is determined by 
\begin{equation}
\sum_{k=1}^4 \vec{a}^{(k)} =0, \;\;\; \mbox{and} \;\;\; 
\vec{a}^{(k)}\cdot \vec{a}^{(k')} = {\rm const} \equiv \cos(\theta).
\end{equation}
This implies $\theta =\cos^{-1}(-1/3)$ so that for Eq. (\ref{Hz}) 
the set of rotations can be 
$\Bid, R_{\hat{y}}(\theta), R_{\hat{a}_2}(2\pi/3), R_{\hat{a}_2}(-2\pi/3)$ 
acting on the initial vector $\vec{a}_1=(0,0,g)$, and $\hat{a}_2$ is
the direction of $\vec{a}_2=R_{\hat{y}}(\theta)\vec{a}_1$.
These rotations 
will take $\vec{a}$ to different positions which 
correspond to the vertices of a tetrahedron 
(see Fig. \ref{tverts}).  The corresponding $U_k$ are found 
using Eq. (\ref{Rn-un}).

Note that this last example uses a set of rotations that 
does not satisfy the subgroup condition. This shows 
that the subgroup condition is sufficient, but not necessary.  
Though it is not a necessary condition, it is 
important due to its convenience.  
The conditions both necessary and sufficient 
for first order decoupling are that the sum of the modified 
Hamiltonians, defined by the modified adjoint vectors, 
sum to the desired effective Hamiltonian.


\subsection{Two Qubits}

\label{two-qubits}

\begin{center}
\begin{figure}[hththt]
\unitlength1mm
\begin{picture}(80,55)
\put(10,0){
\resizebox{4.5cm}{!}{\includegraphics{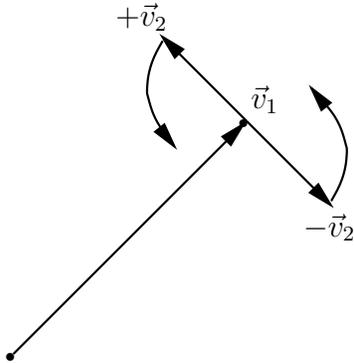}}}
\put(11,0){\circle*{1}}
\put(43,33){{\large{$\vec{v}_1$}}}
\put(50,16){{\large{$-\vec{v}_2$}}}
\put(25,44){{\large{$+\vec{v}_2$}}}
\put(42,31){\circle*{1}}
\end{picture}

\caption{Two qubit computation.}
\label{2q}

\end{figure}
\end{center}

Now let us suppose a computation is to be performed on two-qubits 
using the Heisenberg exchange coupling in the presence of a 
collective dephasing mechanism.  From the study of computation on 
decoherence free subspaces (DFS), we know that this is possible since 
the Heisenberg interaction commutes with the group elements 
that form the stabilizer of the DFS \cite{Kempe:00}.  Our goal here
will be to interpret this condition geometrically.

Let a basis for the Lie algebra be given by 
\begin{equation}
\{\lambda_i\}_{i=0}^{15} = \{ \sigma_i \otimes \sigma_j\}_{i,j=0}^{3},
\end{equation}
and labels correspond to 
\begin{equation}
\begin{array}{lcl}
\lambda_j,\;\; j = 0,1,2,3 &\;\;\;\leftrightarrow\;\;\;& \sigma_i \otimes \Bid, \;\;i=0,1,2,3, 
\\
\lambda_j, \;\;j = 4,5,6&\;\;\; \leftrightarrow \;\;\;&\Bid\otimes \sigma_i,\;\; i=1,2,3,
\\
\lambda_j,\;\; j = 7,8,9 &\;\;\;\leftrightarrow\;\;\;& \sigma_1\otimes \sigma_i,\; i=1,2,3,
\\
\lambda_j, \;\;j = 10,11,12 &\;\;\;\leftrightarrow \;\;\;&\sigma_2\otimes \sigma_i,\; i=1,2,3,\;\;\;
\\
\lambda_j,\;\; j = 13,14,15& \;\;\;\leftrightarrow \;\;\;&\sigma_3\otimes \sigma_i,\;i=1,2,3.
\end{array}
\end{equation}
This forms an orthogonal basis with respect to the trace and has 
normalization given by 
\begin{equation}
\mbox{Tr}(\lambda_i\lambda_j) = 4\delta_{ij}.
\end{equation}

The Heisenberg interaction can be written as:
\begin{equation}
H_{ex} = J\vec{\sigma}_1 \cdot \vec{\sigma}_2
         \equiv \vec{v}_1\cdot \vec{\lambda},
\end{equation}
where $\vec{v}_1=(0,0,0,0,0,0,J,0,0,0,J,0,0,0,J)$, so that 
\begin{eqnarray}
H_{ex} = J(\lambda_{7}+\lambda_{11}+\lambda_{15})
\end{eqnarray}
The collective dephasing is given by:
\begin{equation}
H_{I} = g(\sigma_3 \otimes \Bid + \Bid\otimes \sigma_3)\otimes B 
      \equiv \vec{v}_2\cdot \vec{\lambda} \otimes B \nonumber.
\end{equation}
where $\vec{v}_2 = (0,0,g,0,0,g,0,0,0,0,0,0,0,0,0)$.  So 
\begin{equation}
H_{I} = g(\lambda_3+\lambda_6)\otimes B.
\end{equation}
Note that ${\rm Tr}(H_{ex} H_I) = 4\vec{v}_1 \cdot \vec{v}_2 = 0$ so
$\vec{v}_1 \perp \vec{v}_2$.  A method for achieving 
the effective decoupling without the loss of the desired 
Heisenberg interaction is to consider the little group of $\vec{v}_1$.
(The set of rotations that leaves this vector fixed.)  
From that set of rotations, a subset of rotations exists which 
will rotate the interaction Hamiltonian since the two 
vectors lie in orthogonal subspaces. These rotations clearly must be about the axis defined by $\vec{v}_1$.  Thus we may express this as 
$R_{\hat{v}_1}(\theta)$.  To limit the number of 
pulses in the sequence of BB operations, a parity-kick 
operation is desired.  This further limits our choices to 
those operations that rotate $\vec{v}_2$ by an angle $\pi$.  More specifically, we seek a rotation that
inverts the components in the directions $\lambda_3$ and $\lambda_6$,
since $\vec{v}_2 = g(\lambda_3+\lambda_6)$. The directions $\lambda_3$ and
$\lambda_6$ define a plane perpendicular to $\hat{v}_1$, so that the
desired rotation matrix is effectively an $SO(3)$ rotation matrix with
a non-trivial component in this plane, i.e., $R_{\hat{v}_1}(\pi)$. It
is then simple to check that the corresponding unitary operation
satisfying the parity kick condition $U^\dagger H_{I} U =
R_{\hat{v}_1}(\pi) H_{I} = -H_{I}$, is (see Fig. \ref{2q}) 
\begin{equation}
U \equiv U_1 U_2 = \exp(-i(\sigma_1^{(1)}+\sigma^{(2)}_1)\pi/2)
         = -\sigma^{(1)}_1\sigma^{(2)}_1,  
\end{equation}
where the superscript indicates the qubit on which the 
operator acts.  This interaction leaves $H_{ex}$ unaffected and provides decoupling 
equivalent to Eq.~(\ref{Hgeoeffsu2}).
Note that 
this is a useful means for achieving the 
desired decoupling, because 
exchange interactions can be turned on during a gate 
operation in a solid state device and the decoupling 
can be achieved during the process without interruption of 
the desired interaction. The geometric picture shows that the above
$U$ is by no means unique: any discrete $SO(3)$ subgroup acting in the
$(\lambda_3,\lambda_6)$ plane, and whose
elements add up to zero, will do.


\section{Conclusion}

A geometric treatment of bang-bang (BB) operations has been provided.  
This perspective provides an intuitive picture for BB operations and 
their imperfections.  The group averaging is made explicit 
through the corresponding average over a set of coordinate vectors
representing rotations of the Hamiltonian; the 
resultant vector is the sum over all the BB modified Hamiltonians.  
These quantities are useful 
for computations, complementing the somewhat more abstract 
approaches of previous treatments \cite{Viola:00a,Zanardi:99d}.  
Since after the application of 
an imperfect set of decoupling operations, one is concerned 
with remaining error(s), such tools are useful for visualization.  
The often promoted Bloch sphere representation used in some of the 
examples treated here provides a means for extending intuition beyond the 
the low dimensional cases.  

The usual group-theoretic symmetrization description of BB operations 
assumes that the set of pulses forms a discrete subgroup
\cite{Viola:00a,Zanardi:99d}. We showed here that this is not a
necessary condition, through the example of symmetrizing by the
vertices of a tetrahedron.  This fact has been well appreciated 
in the context of recoupling schemes in NMR 
(see for example \cite{Ernst:book}).  Here we wish to 
emphasize it in the 
context of quantum information processing.

The two-qubit example in Section \ref{two-qubits} provides a way in 
which this geometric analysis aids in the problem of 
finding decoupling interactions.  The similarity between 
this example and recoupling techniques in NMR and 
other systems is no coincidence.  The BB operations 
were, after all, related to NMR techniques in the earliest 
papers describing such interactions for quantum error 
correction \cite{Viola:98}.  
The geometric viewpoint is quite general and provides an 
instructive way in which to decompose such problems.  They 
may be particularly useful for the types of recoupling 
techniques one requires for reducing constraints on 
quantum computer proposals \cite{Lidar/Wu}.

In the subgroup framework, our geometric
picture uses a homomorphic mapping
between the  fundamental representation 
and the corresponding  adjoint representation.  
The problem of inverting this map from 
the adjoint to the fundamental representation may well be 
difficult for groups of higher dimension than $SU(4)$.  
However, for universal 
quantum computation, one and two qubit gates are sufficient and 
fortunately the discrete subgroups of unitary groups have 
been classified up to $SU(4)$. (See 
\cite{Hanany/He,Fairbanks/Fulton/Klink,Anselmi/etal} 
and references therein).  The determination of the 
appropriate subgroup could consist of searching a 
discrete solution space. This appears feasible since 
the lower order subgroups are 
more relevant given the strict time constraints of 
the BB assumptions.

Given the scarcity of qubits in current quantum computing
systems, we believe that the BB method is an important tool. 
We hope that the work presented here will be helpful in 
constructing sequences of
BB pulses and analyzing their imperfections.


\begin{acknowledgments}
 DAL acknowledges support from PREA, PRO, the
Connaught Fund, and AFOSR (F49620-01-1-0468).  MSB acknowledges 
S. Schneider and K. Khodjasteh for help creating the graphics.  
We would also like to thank Lorenza Viola for several useful 
comments.
\end{acknowledgments}



\end{document}